# Machine learning for violence prediction: a systematic review and critical appraisal

Stefaniya Kozhevnikova[1], Denis Yukhnenko[1], Giulio Scola[1], Seena Fazel[1,2].

1 – Department of Psychiatry, University of Oxford, Oxford, United Kingdom
2 – Oxford Health NHS Foundation Trust, Oxford, United Kingdom

Abstract

Purpose

To conduct a systematic review of machine learning models for predicting violent behaviour by synthesising and appraising their validity, usefulness, and performance.

Methods

We systematically searched nine bibliographic databases and Google Scholar up to September 2025 for development and/or validation studies on machine learning methods for predicting all forms of violent behaviour. We synthesised the results by summarising discrimination and calibration performance statistics and evaluated study quality by examining risk of bias and clinical utility.

Results

We identified 38 studies reporting the development and validation of 40 models. Most studies reported Area Under the Curve (AUC) as the discrimination statistic with a range of 0.68-0.99. Only eight studies reported calibration performance, and three studies reported external validation. 31 studies had a high risk of bias, mainly in the analysis domain, and three studies had low risk of bias. The overall clinical utility of violence prediction models is poor, as indicated by risks of overfitting due to small samples, lack of transparent reporting, and low generalisability.

Conclusion

Although black box machine learning models currently have limited applicability in clinical settings, they may show promise for identifying high-risk individuals. We recommend five key considerations for violence prediction modelling: (i) ensuring methodological quality (e.g. following guidelines) and interdisciplinary collaborations; (ii) using black box algorithms only for highly complex data; (iii) incorporating dynamic predictions to allow for risk monitoring; (iv) developing more trustworthy algorithms using explainable methods; and (v) applying causal machine learning approaches where appropriate.

# Introduction

Violent behaviour is an important challenge for public health. Despite violence being a rare outcome in people with mental disorders (less than 10% in people with personality disorders and schizophrenia spectrum disorders, and less than 5% in people with other disorders(1)), violence prevention is essential in these populations to support recovery. Exhibiting violent behaviour in a medical facility, or possible arrest, can disrupt care and result in the loss of a support system, consequently reducing the effectiveness of a treatment programme. Furthermore, in some areas, mostly low- and lower-middle-income countries, criminal violence is one of the leading causes of death or serious injury (2, 3). Even in higher-income countries, such as the USA, domestic violence is a leading cause of maternal mortality (4). These issues create a high demand for violence prevention systems in healthcare and criminal justice settings. One component of a functional and effective prevention system is violence prediction, which will inform specialists' decisions.

Different violence prediction tools have existed for decades (5). In the recent years, the development of decision-aiding tools has shifted from simple regression-based modelling to the use of more advanced statistical methods, known as machine learning. "Machine learning methods", or "artificial intelligence", is an umbrella term describing several computational methods that perform a task based on a previously existing (training) dataset (6, 7). In other words, applying machine learning methods means creating an algorithm that "learns" patterns and rules from the data and applies these rules to new input data, without directly programming the task-solving process (8). Many different methods exist in this field, and the list continues to expand as artificial intelligence develops rapidly.

One of the ways to describe machine learning methods is roughly dividing them into "interpretable" and "black box" methods. Interpretable machine learning methods are algorithms with established, direct connections between predictors and outcomes, such as Naïve Bayesian (9) or decision trees (10). Black box methods, on the other hand, are typically more complex algorithms, such as neural networks (11) or random forests (12). While interpretable methods are intuitively understandable for humans, the internal processes of black box methods are not, and understanding them requires additional training in either the specific method or advanced statistics more generally. Black box methods often achieve higher performance than traditional methods (13), including in healthcare (14). However, experts note that inability to understand and assess their decision-making process creates a lack of trust among specialists. As stated in a review on cardiological predictions (15), the clinical utility of black box models in healthcare remains questionable. The authors emphasise the importance of understanding the causes underlying model predictions to avoid unnecessary investigations, particularly in cases of false positives. A review of clinical prediction models in psychiatry comes to the same conclusion: despite rapid advances in the field, the clinical utility of existing models remains highly limited (16). In this review we aimed to assess the perspectives of using black box methods in the field of violence prognosis: is the superior performance of black box algorithms useful in violence prediction?

To assess the clinical utility of prognostic models, evaluating the models separately from their development process is insufficient. As the field has progressed, standards have been established to ensure high-quality prediction models. High-quality models are neither under- nor overfitted. An underfitted model does not produce sufficiently accurate predictions from the data. Overfitting, on the other hand, occurs when a model uses too much information from the training data, capturing noise and idiosyncrasies instead of estimating new outcomes (11). An overfitted model produces outputs similar to the training data, failing to generalise patterns beyond the training set, and therefore has limited practical value. Current standards cover various stages of model development, including sampling procedures and data analysis, to ensure that models perform well. It is also recommended to employ different performance measures—most notably to assess both a model's ability to correctly distinguish between positive and negative outcomes (discrimination) and the closeness of predicted outcomes to actual observations (calibration) (17, 18).

The purpose of this review is to synthesise data on black box models for violence prediction and to appraise their validity, usefulness, and performance. Our objectives were to thoroughly examine the field of machine learning for violence prediction, assess model performance for different types of outcomes, critically evaluate clinical utility, and outline the limitations and prospects for developing such tools.

## Methods

### Protocol and registration

The PROSPERO protocol for this review has been prepared before carrying out literature search in July 2024 and is available in Supplementary materials (Supplementary A).

### Literature search

We followed the Preferred Reporting Items for Systematic Reviews and Meta-analyses guidelines (PRISMA) (19), and Transparent Reporting of Multivariable Prediction Models for Individual Prognosis or Diagnosis: checklist for systematic reviews and meta-analyses (TRIPOD-SRMA) (20). We searched several databases: Ovid (MEDLINE, Embase, APAPsychArticles, AMED, BIOSIS, Global Health, PsycINFO), dissertation databases, and Web of Science Core Collection to identify the papers reporting the development and/or validation of models predicting any type of violent behaviour (i.e. violent behaviour in general, violent crime and others. On the later stage we also manually searched Google Scholar and grey literature to identify potentially missed studies. No publication date or language restrictions were set. The initial search was conducted in July 2024 and updated in September 2025. The same keywords were used for all databases.

*((violen\* or recidiv\* or crim\* or reoffen\*) and predict\* and (machine learning or deep learning or AI)).mp. not (systematic review or meta-analysis or meta analysis or narrative review or literature review).ab*

### Eligibility criteria

Studies were considered eligible for inclusion if they described the development, internal and/or external validation of a multivariable black box machine learning model for predicting any violent behaviour in any target audience. For inclusion criteria of outcome definition, we used the World Health Organisation definition: "the intentional use of physical force or power, threatened or actual, against oneself, or against a group or community that either results in or has a high likelihood of resulting in injury, death, psychological harm, maldevelopment or deprivation" (21). We considered for inclusion the following black box methods: neural networks, support vector machines, k-nearest neighbours methods, random forest methods, including gradient boosting, super learners (models using a set of several methods), and natural language processing methods (NLP).

We have excluded papers that do not account for violence as a separate outcome, i.e. papers that focus on recidivism in general. We also excluded papers that predict victimisation of a person, rather than a violent behaviour of them. In addition, we have excluded reviews and theoretical papers and papers that focused on biological variables, such as neuroimaging or genetics, due to pragmatic concerns.

Study selection

In accordance with PRISMA guidelines, two reviewers were involved in the study selection process. At first, SK reviewed the titles and abstracts of all identified studies, and GS double-screened and validated a random sample of 20% titles and abstracts. The same procedure was conducted in assessment of eligible full texts. Any disagreements on study selection were resolved in consultation with SF and DY.

Data extraction

Data extraction was carried out by SK following the Critical Appraisal and Data Extraction for Systematic Reviews of Prediction Modelling Studies (CHARMS) checklist (22) using the automated template (23). Data extraction was later verified by GS. Information on the following variables were collected: (i) sample demographics; (ii) setting in which data were collected, e.g. community, probation, hospital or prison; (iii) study design; (iv) performance estimates of the model, including measures of discrimination and calibration.

When articles reported development and testing of multiple models, for example, comparing neural network and random forest, we included the model with the highest discrimination and calibration measures. If models assessed different outcomes (i.e. domestic violence and general violence) or used different follow-up periods, we included all outcomes and follow-ups as separate models.

Summary measures

To assess the predictive performance of the included models, we extracted measures of discrimination and calibration. The area under the receiver operating characteristic curve (AUC) is currently one of the most widely used measures of discrimination, and we extracted it wherever possible. It expresses the probability that the model will assign a higher risk score to a violent person. Elsewhere, especially in the older studies, we extracted other metrics of model performance, such as accuracy or F1 score. We did not rely on AUC thresholds in our review because such thresholds are chosen post-hoc by the researchers and can lead to overinflated findings REFERENCE. Where possible, we also extracted measures of calibration, such as Brier score or calibration error (CalErr). Since the outcome of interest is binary, we considered root mean square error (RMSE) as calibration measurement, as for binary outcome $RMSE = \sqrt[2]{Brier\ score}$ (24).

Risk of bias and publication bias

The risk of bias within each study was assessed with the Prediction model Risk of Bias Assessment Tool (25). The PROBAST was developed specifically for evaluating studies of diagnostic and prognostic prediction models and provides ratings of the risk of bias in four different domains, namely: (i) participants, (ii) predictors, (iii) outcome, and (iv) analysis. PROBAST domains assess how appropriate were the procedures carried out during model development and validation, such as variables selection, internal and external validation and performance measuring. For each domain there are several guidelines that ensure transparent reporting and minimise the risk of creating under- or overfitting of the model. Important part of the model development that prevents overfitting is using a large enough data set for training, especially ensuring that there is a reasonable ratio of positive outcomes and variables used (EPV). Depending on the model the EPV ratio should be over 10 for regression-based models (26) and over 20 for models using more sophisticated methods (25, 27). Other important procedures to lower risk of bias include using appropriate missing data handling (participants should not be eliminated due to missing data), avoiding data-driven predictors selection, and using both discrimination and calibration performance measures. Based on the established risk of bias, the overall risk is assessed. The overall risk of bias should be considered high if at least one of the domains has shown high risk, and if the model hadn't undergone external validation unless the original model was developed with a very large dataset.

Data synthesis

Due to the high heterogeneity of the reported models, a narrative data synthesis was performed.

Results

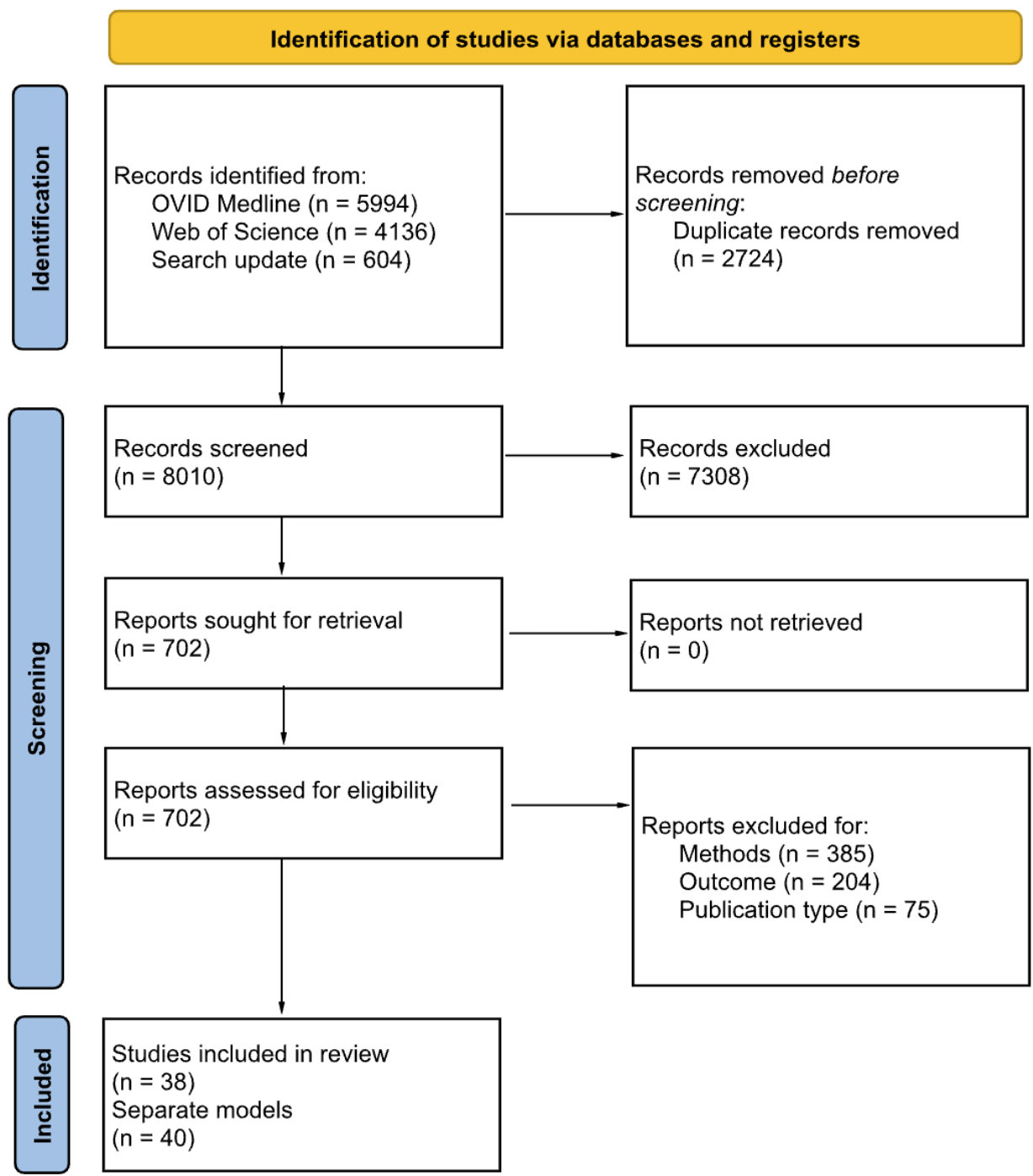

Figure 1. PRISMA flowchart of included studies. Studies excluded for publication type were conference proceedings, reviews and other reports with no data related to the current review.

Characteristics of the included studies are presented in Table 1. Overall included 40 models from 38 studies (Figure 1). We divided included studies into two large groups: criminal violence (26 studies) and other violent behaviour (12 studies). Within these groups we created subgroups depending on the outcome. The total number of participants was 2,362,029 people, most of which were male (87.7%). For the detailed breakdown of all studies included, please refer to the Supplementary materials (Supplementary B).

| Outcome | Number of studies | Total number of participants with outcome / overall | Mean prediction time, weeks (years) | Mean participant age, years | Mean male % |
|---|---|---|---|---|---|
| Violent crime | | | | | |
| General criminal violence | 6 | 6424 / 979428 | 260.7 (5.0) | 34.9 | 90.5 |
| General violent recidivism | 13 | 2679 / 583102 | 195.6 (3.8) | 31.2 | 88.9 |
| Intimate partner violence | 5 | 398 / 496248 * | Not reported | Not reported | 69.7 |
| Intimate partner violence reoffence | 4 | 1492 / 71861 | 156.43 (2.8) | 32.4 | 100 |
| Extremism | 2 | 1420 / 2328 | Not reported | Not reported | Not reported |
| Other violent behaviour | | | | | |
| General violent behaviour | 4 | 5251 / 34460 | 126.01 (2.4) | 32.3 | 88.5 |
| Inpatient violence | 8 | 2314 / 255766 * | 44.9 (0.9) | 38.7 | 67.4 |

Table 1. Features of the included studies divided by the outcomes. * – include one or more studies with natural language processing and/or image recognition, where only the overall number of cases is known.

Within black box methods used, the most common ones are random forests (12 studies, 30%); and there were 9 studies (22.5%) using gradient boosting methods and the same number of neural networks. Three studies (7.5%) reported using super learners – algorithms that consist of several different models. There was only one study that used k-nearest neighbours methods, specifically, the unsupervised k-means. Figure 2 provides a more detailed breakdown of ML methods used.

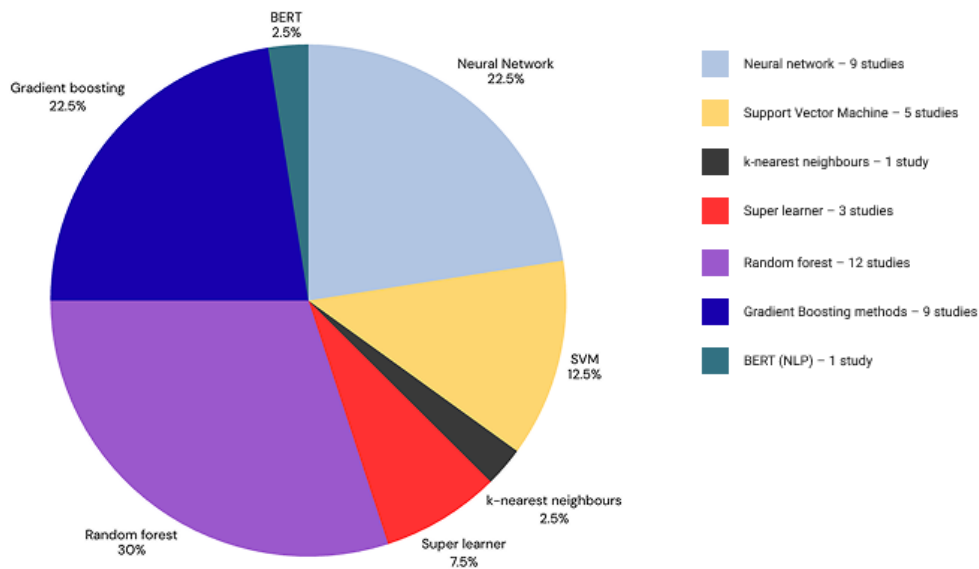

Figure 2. Distribution of methods used in the included papers.

One study used image recognition of a projective psychological test along with other variables, and 5 (13.2%) studies used natural language processing to predict outcomes from police reports or medical records. One study (28) used a model based on the Bidirectional Encoder Representations from Transformers (BERT), an NLP model previously developed and published by Google (29).

29 out of 38 (71.8%) studies reported area under the curve (AUC) as an indication of performance. Other papers used accuracy and F1-score as discrimination metric. Eight studies (21%) reported calibration performance, one paper reported Brier score, two papers reported RMSE and calibration error, four studies used calibration plots, and two studies used Brier score or RMSE but did not report it. All studies reported internal validation, mostly using k-fold cross-validation, and only three studies reported conducting external validation (Figure 3).

Out of 37 studies that were not doctorate dissertations, only six studies (15.8%) had both mental health or judiciary specialist and a computer scientist as either a first, second or a senior author. Nine studies were led by computer scientists, and 22 – by either a mental health or judiciary specialists.

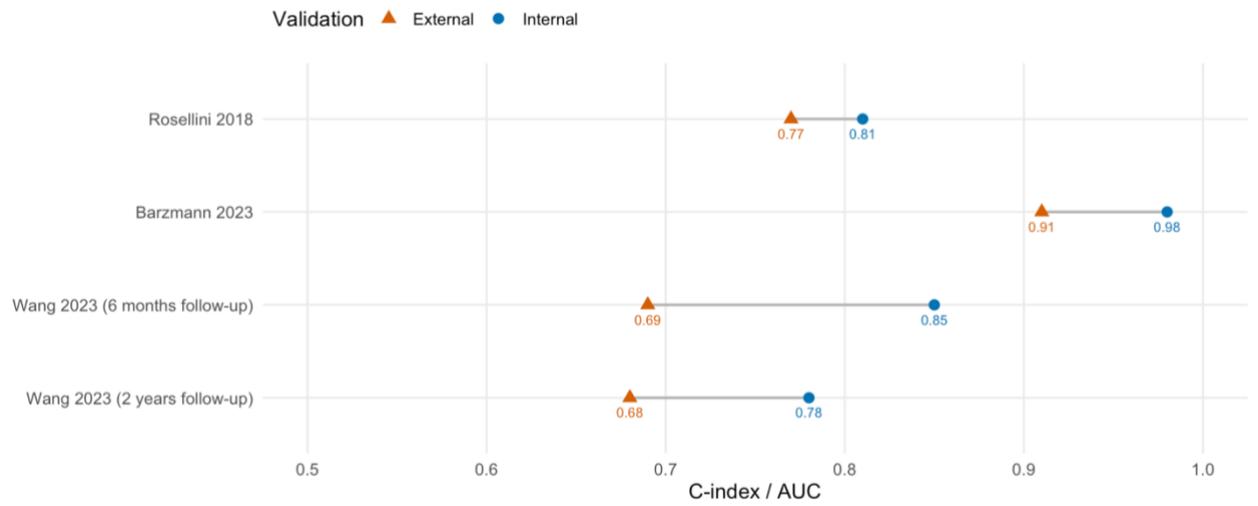

Figure 3. Discrimination metrics in the studies reporting external validation. Internal AUC reflects model performance on the training dataset, while external AUC reflects performance on a different dataset (e.g., another site or time period). AUC ranges from 0.5 (no discrimination) to 1.0 (perfect discrimination).

Most of the models were developed on populations from high- or middle-income countries with the high-income countries being prevalent in the number of participants (Figure 4).

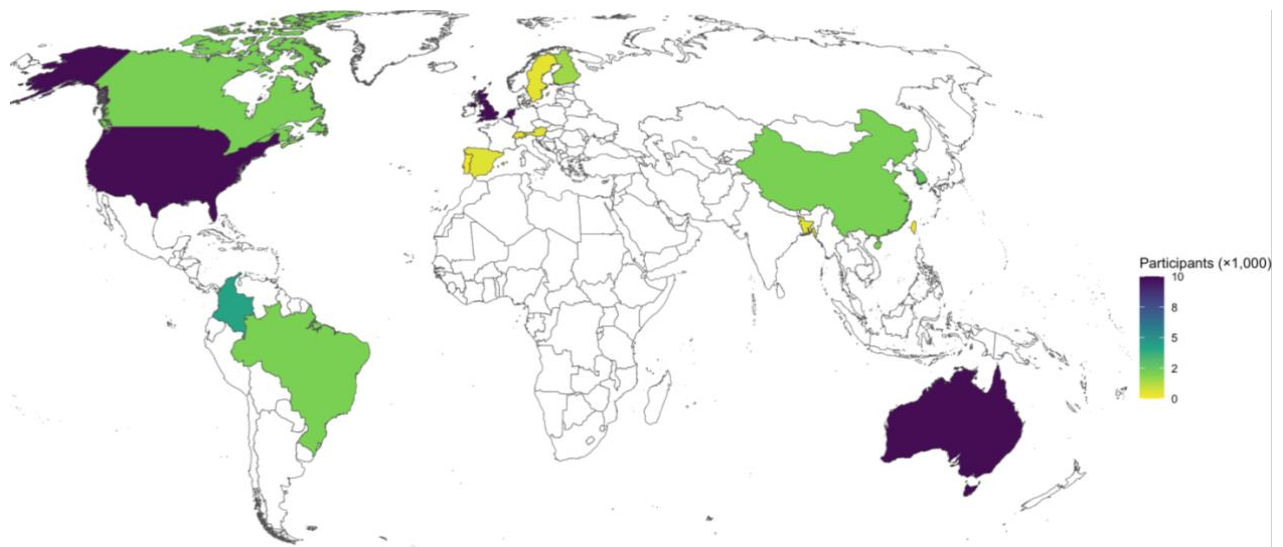

Figure 4. Geographic coverage of study samples in participant numbers in thousands.

Risk of bias

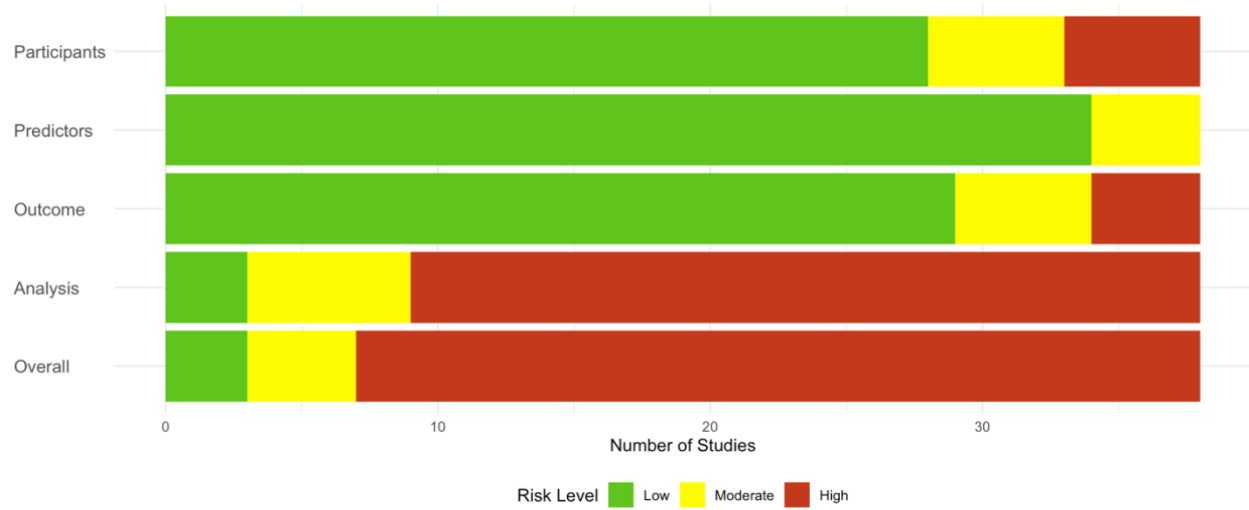

Figure 5. Risk of bias in the included studies

The risk of bias was assessed with the PROBAST tool. Based on the PROBAST guidelines, we established the overall the risk of bias in the included studies high in 31 papers (81.6%). The highest risk of bias in the included papers was related to the analysis domain (Figure 5). Below we briefly outline the most common risk factors across all studies.

The first common risk factor for bias lies in the domain of participants' selection. It is important not to oversample participants with positive outcome and ensure that the high-risk participants are not prevalent in the training dataset. However, eight studies (21%) did not provide any risk assessment and/or adjustment for included cohorts, while the sampling procedures were not described and may have not been randomised. This may have created an oversampling of participants with positive outcome.

Most of the models were developed using the data from retrospective cohort studies, which has created risks of bias in the domain of outcome definition and assessment. In six studies (15.8%) there is a risk that predictor variables could be a part of outcome definition, and for five studies (13.2%) it is unclear, whether outcomes were determined without knowledge of predictors data.

As mentioned above, the highest risk of bias was established in the analysis domain. Since we have focused on the black box algorithms, events per variable ratio should be over 20, as recommended per PROBAST guidelines. Only six models (15.8%) from five studies had EPV over 20, and five models (12.5%) had EPV over 10, which would have been acceptable for regression-based prediction models. For 9 models it was impossible to establish EPV, and other 20 models (50%) had EPV less than 10, which means that over half of the published prediction models are at risk of overfitting due to very low EPV, while there were no additional preventative measures taken, like regularisation techniques. Calibration performance was measured for nine models (22.5%) but not reported for two of them. There is no evidence of calibration performance being carried out for other models. Moreover, despite data-driven approach being recommended against for predictive modelling, 8 models (21%) used univariable analysis for the initial prediction selection.

Overall, three studies reporting five models showed low risk of bias, and only one study reporting two models for recidivism prediction with different follow-up periods (30) carried out all required procedures to minimise the risk of bias. The reported models were based on SVM and Gradient Boosting methods, and the studies were primarily led by computer scientists.

## Discussion

This review summarises over 30 years of research in violence-prediction modelling using black box machine-learning algorithms. It provides a comprehensive overview of 40 models, with most studies focused on model development. We included 38 studies covering over 2 million participants. There is clear evidence of exponential growth in the field, especially since 2018, and an increasing use of deep-learning algorithms, including natural language processing. Between the first literature search in July 2024 and the search update in September 2025, six new predictive models were published, three of which were black box models eligible for inclusion in this review. Such rapid progression is relatively rare in other fields and raises concerns regarding the quality and clinical utility of these models.

In general, published models showed good discrimination performance, but reporting calibration performance was scarce, despite the importance of this metric in prediction modelling (31). The common issue was incomplete reporting compared with the TRIPOD+AI guidelines (32). The PROBAST assessment also highlighted challenges and areas for improvement, particularly in the analysis domain. A major issue currently lies in methodological shortcomings that may lead to overfitting: datasets are often not large enough for sufficient model training; some models are based on retrospective data without clarifying the temporal separation between predictors and outcomes; and variable selection is sometimes based on data-driven approaches. External validation was conducted in only three studies, and one of these (30) repirted limited generalisability of the developed models.

Based on the results of this review, we propose five key considerations for future research on violence-prediction modelling. First, a higher level of adherence to quality standards and stronger interdisciplinary collaboration should be ensured. We recommend that researchers routinely use quality-assessment tools such as PROBAST to self-evaluate risk of bias, and follow reporting guidelines more thoroughly. We also observed variation in interdisciplinary collaboration across author teams. Currently, a minority of studies are co-authored by both clinicians and computer scientists. In studies led mainly by clinicians, analytic methods were at greater risk of bias. We noted occasional imprecision in terminology (e.g., mixing "classification" with "prediction") and limited discussion of assumptions and other important features of complex algorithms. Higher-quality studies led by computer scientists often did not aim to solve clinical problems but instead explored the potential of specific methods. Greater collaboration between mental-health and judiciary specialists and computer scientists would help ensure that well-trained models address high-priority issues. Therefore, our first recommendation is routine self-assessment of model quality and equal involvement of mental-health or criminal-justice experts and computer scientists in model development.

Second, most of the data used in published models were routine administrative and/or clinical data without complex interactions or high-dimensional features. One key principle of clinical prediction modelling is to use simpler methods (e.g., Cox proportional hazards models, logistic regression) for structured data with a moderate number of predictors (33). The use of complex black box models for modelling on such datasets is unlikely to lead to superior results compared with classic models (34), a finding supported by one of the studies in this review (35). Moreover, logistic regression models usually calibrate better on routine data (36) and provide more stable results (37). We therefore recommend using black box models only on large, complex datasets.

One way to improve predictive accuracy in current models is to incorporate temporal proximity, where possible. Risk of violence is dynamic and influenced by multiple aspects of an individual's life. For example, the risk of reoffending decays over time and, after several years, falls to a level comparable to that of the general population (38). Studies also suggest that different factors affect violence risk over different time frames, with some factors associated with heightened risk within the first week after occurrence (39). Thus, scores need to be updated frequently to provide optimal accuracy (40). Temporal proximity and dynamic predictions are commonly implemented in psychiatry using Cox proportional hazards regression with sliding window landmarks (41), and recent studies on dynamic prediction with black box models in healthcare have shown promising results (42-44). If data allow dynamic prediction, we recommend using this approach to incorporate risk monitoring.

As noted earlier, limited understanding of how models reach their decisions creates mistrust and hesitancy to implement even high-performing algorithms with low risk of bias. In clinical settings, risk estimates are most valuable when linked to modifiable factors. In the judiciary system, transparency in decision-making is essential and widely debated, including in relation to limiting systemic biases and ensuring fairness (45). Therefore, our next recommendations focus on methods that increase transparency.

One approach to enhance transparency in black box algorithms is the use of explainable AI (XAI) (46), which introduces additional steps to highlight the reasons behind model decisions. This approach has been criticised by some AI in healthcare specialists for either overcomplicating implementation (47) or for justifying unnecessary use of black box algorithms (48). It is important to note that such methods do not establish the causal link between variables and outcomes (15). However, they can help build trust among practitioners and facilitate implementation. Developing mental-health–related models using approaches such as the Transparency and Interpretability Framework for Understandability (TIFU) (49) may ensure more transparent and trustworthy systems.

Another approach is causal machine learning, where the algorithm builds a structural causal model (50). he main benefit of this method is its ability to estimate personalised predictions, leading to higher applicability. Although the utility of causal machine learning for violence prediction remains to be studied, it shows promise in other healthcare-related predictions (51). Causality and its implementation in mental health-related prediction is currently a major topic of debate (52, 53). Not all decisions predicted by algorithms require a transparent causal link: accurate indication alone may support effective allocation of resources for triage and assessment. However, in situations requiring causal inference, and where data permit it, we recommend the use of causal machine learning.

All the recommendations above require a comprehensive understanding of both machine-learning algorithms and the field of application. Ensuring that all relevant features of the data are considered, that models are developed and validated according to high-quality standards, and that predictions benefit healthcare and judiciary practitioners, patients, and others involved in the justice system requires strong interdisciplinary collaboration. Therefore, our first recommendation, enhanced collaboration, remains the most fundamental.

This study provides a comprehensive overview of existing machine learning models developed for violence prediction. Recent reviews have either focused on discrimination performance (54), or on to specific populations and tool evaluations (55, 56). To our knowledge, this is the first attempt to synthesise all available evidence on predicting any form of violent behaviour using black box models. However, several limitations should be noted. As there are limited means to systematically search grey literature, some relevant data may have been overlooked, despite our use of broad search terms and additional manual searching. Included studies were highly heterogeneous, precluding meta-analysis.

## Conclusion

Despite rapid progress in the use of machine learning for violence prediction, methodological shortcomings, limited geographic coverage, and the lack of external validation mean that the predictive performance of these tools is currently uncertain, and, on the basis of this systematic review, we do not recommend any current model for implementation. The best approach to prediction modelling in medical-related fields is to assess every methodological decision (e.g., methods to use, sample size, feature selection) on a case-by-case basis, considering the overall impact of any potential feature of a model. We outline five key areas to improve research quality and potential clinical utility: (i) ensure higher methodological quality of model development and validation, which involves collaboration between mental health and computer science/AI/data science experts; (ii) use black box algorithms only with highly complex data, while retaining reliable traditional methods for routine, low-dimensional data; (iii) where possible, incorporate temporal proximity and dynamic prediction approaches to allow for risk monitoring; (iv) when black box models are developed, incorporate more trustworthy algorithms using XAI; and (v) in predictions requiring a transparent causal link, apply causal machine-learning methods.

## Contributors

SK completed data screening, data extraction, data curation, data visualisation, and writing (original draft and editing). GS conducted the abstract and title screening and reviewed data extraction. DY and SF led on conceptualisation, supervision, and writing (review and editing). SK and GS directly accessed and verified the underlying data reported in the manuscript. All authors had full access to all the data in the study and accept responsibility to submit for publication.

**Supplementary materials**

Supplementary A. PROSPERO protocol

Review title: Machine learning for predicting violent behaviour: a systematic review.

Condition or domain being studied: Violent behaviour; Violence; Violence Prevention; Crime; Machine learning

Rationale for the review: There is a growing number of machine learning prediction tools for violence that are being developed, and little is known about the quality of such tools.

Review objectives: What is the quality and clinical applicability of machine learning models that predict violent behaviour?

Keywords: Machine learning; Violence; Prediction

Country: United Kingdom

Eligibility criteria:

- Population: General population, prison population, psychiatric populations. No exclusion criteria was set for the population.
- Intervention(s) or exposure(s): This review will be exploring the predictive modelling tools designed using machine learning methods, such as neural networks, kNN, random forest etc. Excluded: white box and non-machine learning models.
- Comparator(s) or control(s): People who do not exhibit violent behaviour. No exclusion criteria was set.
- Study design: Only nonrandomized study types will be included. Observational (retrospective and prospective cohort) studies that resulted in development and/or validation of the prediction model will be included.

Searching and screening:

- Both published and unpublished studies will be sought.

- The main databases to be searched are Embase - Embase via Ovid, MEDLINE, PsycInfo, PubMed, SCI - Science Citation Index and Scopus.
- Other important or specialist databases that will be searched: Dissertations, arxiv.org, Google Scholar.
- There are no language restrictions.
- There are no search date restrictions.
- Other studies will be identified by: contacting authors or experts, reference list checking, searching conference proceedings and searching dissertation and thesis databases.
- Studies will be screened independently by at least two people (or person/machine combination) with a process to resolve differences.

Data extraction from published articles and reports:

- Data will be extracted independently by at least two people (or person/machine combination) with a process to resolve differences.
- Authors will be asked to provide any required data not available in published reports.
- Risk of bias will be assessed using: *PROBAST*
- Data will be assessed by one person (or a machine) and checked by at least one other person (or machine).
- Additional information will be sought from study investigators if required information is unclear or unavailable in the study publications/reports.

Outcomes to be analysed: Any violent behaviour: violent recidivism, terrorism, murder, domestic violence etc. There are no additional outcomes.

Strategy for data synthesis: Narrative synthesis is planned for this review. We are planning to extract and synthesise all data related to model performance, internal or external validation and calibration of the models analysed.

Supplementary B. All included studies

| First author, year | Country | Model | Number of outcomes / number of variables; (EPV) | Discrimination metric | Calibration metric | External validation (performance) |
|---|---|---|---|---|---|---|
| General criminal violence | | | | | | |
| Gordon, 1992 | USA | Backpropagation neural network | 106 / 17; (6.3) | Recall = 0.76 | No | No |
| Bauto, 2023 | Portugal | Multilayer perceptron | 59 / 9; (6.6) | Accuracy = 0.83 | No | No |
| Rosellini, 2015 | USA | Random forest | 5771 / 446; (12.9) | AUC = 0.81 | No | Temporal (AUC = 0.77) |
| Sonnweber, 2021 | Switzerland | Stochastic gradient boosting | 294 / 10; (29.4) | AUC = 0.76 (95% CI 0.69-0.83) | No | No |
| Watts, 2021 | Canada | XGBoost | 863 / 156 (5.5) | Accuracy = 0.57 | No | No |
| Santos jr, 2023 | Brazil | XGBoost | Not reported / 9 | AUC = 0.99 | No | No |
| General violent recidivism | | | | | | |
| Grann, 2007 | Sweden | Backpropagation neural network | 91 / 10; (9.1) | AUC = 0.72 (95% CI 0.65-0.79) | No | No |
| Liu, 2011 | UK | Multilayer perceptron | 343 / 20; (17.1) | AUC = 0.71 (95% CI 0.66-0.75) | No | No |

| Hamilton, 2014 | USA | Neural network | Not reported / 23 | AUC = 0.73 (SD 0.03) | RMSE = 0.21 CalErr = 0.05 | No |
| --- | --- | --- | --- | --- | --- | --- |
| Pelham, 2020 | USA | Neural network | Not reported / 43 | AUC = 0.78 (95% CI 0.71-0.84) | Brier score = 0.17 (95% CI 0.14-0.19) | No |
| Neuilly, 2011 | USA | Random forest | 168 / 60; (2.8) | Accuracy = 0.58 F1 score = 0.62 | No | No |
| Berk, 2014 | USA | Random forest | 1283 / not reported | Accuracy = 0.73 | No | No |
| Salo, 2019 | Finland | Random forest | 278 / 52; (5.3) | AUC = 0.77 (95% CI 0.70-0.84) | Calibration plot | No |
| Etzler, 2024 | Austria | Random forest | 46 / 37; (1.2) | AUC = 0.78 (SD 0.06) | No | No |
| Zeng, 2017 | USA | Stochastic gradient boosting | 6455 / 48; (134.5) | AUC = 0.72 | Calibration plot | No |
| Laqueuer, 2024 | USA | Super learner | Not reported / 91 | AUC = 0.72 | Calibration plot | No |
| Tollenaar, 2013 | Netherlands | Support vector machine | 380 / 6; (63.3) | AUC = 0.73 | RMSE = 0.40 CalErr = 0.06 | No |
| Wang, 2023 | USA | XGBoost (6-months follow-up) | 1755 / 41; (42.8) | AUC = 0.85 | Calibration plot | Geographic (AUC = 0.69) |
| | | XGBoost (2-year follow-up) | 8526 / 41; (207.9) | AUC = 0.78 | Calibration plot | Geographic (AUC = 0.68) |
| Intimate partner violence | | | | | | |
| Karystianis, 2021 | Australia | Recurrent neural network; NLP | Not reported | AUC = 0.65 | No | No |
| Hossain, 2021 | Bangladesh | Random forest | Not reported / 9 | Accuracy = 0.71 F1 score = 0.76 | No | No |
| Garcia-Vergara, 2023 | Spain | Random forest | 161 / 33; (4.9) | F1 score = 0.87 (SD 0.03) | No | No |
| Verrey, 2023 | UK | Super learner | 237 / 13; (18.2) | AUC = 0.71 | No | No |
| Petering, 2018 | USA | Support vector machine | Not reported / 26 | AUC = 0.69 | No | No |
| Intimate partner violence reoffence | | | | | | |
| Zeng, 2017 | USA | Stochastic gradient boosting | 1183 / 48; (24.6) | AUC = 0.68 | Calibration plot | No |
| Berk; 2020 | USA | Stochastic gradient boosting | 144 / not reported | Accuracy = 0.66 | No | No |
| Extremism | | | | | | |
| Ahmed, 2023 | USA | k-nearest neighbours | 180 / 28; (6.4) | Accuracy = 0.91 | No | No |
| Ivaskevics, 2022 | USA | XGBoost | 1240 / 79; (15.7) | AUC = 0.87 (SD 0.03) | No | No |
| General violent behaviour | | | | | | |

| Santamaria-Garcia, 2021 | Colombia | Neural network | 2117 / 162; (13.1) | AUC = 0.96 | Brier score (not reported) | No |
| --- | --- | --- | --- | --- | --- | --- |
| Rosellini; 2018 | USA | Super learner | 829 / 273; (3.0) | AUC = 0.75 | No | No |
| Brazmann; 2023 | USA | Support vector machine; NLP | 54 / 21; (2.6) | AUC = 0.98 | No | Geographic (AUC = 0.91) |
| Kim, 2023 | South Korea | Support vector machine; image recognition | 134 / not reported | F1-score = 0.98 | No | No |
| Inpatient violence | | | | | | |
| Dobbins, 2024 | USA | BERT; NLP | 246 / not reported | AUC = 0.78 | No | No |
| Menger, 2018 | Netherlands | Recurrent neural network; NLP | 1248 / not reported | AUC = 0.79 (SD 0.02) | No | No |
| Bokhari; 2018 | USA | Random forest | 176 / 31; (5.7) | AUC = 0.75 | No | No |
| Wang; 2020 | Canada | Random forest | 103 / 28 (3.7) | AUC = 0.63 (SD 0.00) | No | No |
| Cheng; 2023 | China | Random forest | 611 / 19; (32.2) | AUC = 0.96 (95% CI 0.94-0.97) | No | No |
| Watts; 2023 | Canada | Random forest | 26 / 67; (0.4) | AUC = 0.91 (95% CI 0.87-0.95) | No | No |
| Hu; 2023 | Taiwan | Random forest; NLP | 406 / not reported | AUC = 0.69 | No | No |

| Le; 2018 | Australia | Support vector machine; NLP | Not reported | Accuracy range (0.69-0.77) | RMSE (not reported) | No |